\documentstyle[12pt]{article}
\begin{document}
%
\large
\thispagestyle{empty}
\vspace*{1cm}
\begin{center}
{\Huge\bf Electromagnetic Fields \\[4mm]
in a Thermal Background}
\\[1cm]
\end{center}
\vspace*{1cm}
\begin{center}
\
{\centering
{\large Per Elmfors,\footnote{Email address:
elmfors@surya11.cern.ch.}}\raisebox{1ex}{,a}
  and
{\large Bo-Sture Skagerstam\footnote{Email address:
tfebss@fy.chalmers.se. Research supported by the Swedish National
Research Council under contract no. 8244-311.}}\raisebox{1ex}{,b}
 \\
{\normalsize\sl \raisebox{1ex}{a}CERN,
TH-Division,
 CH-1211 Geneva 23, Switzerland \\ }
{\normalsize\sl \raisebox{1ex}{b}Institute of Theoretical Physics,
   Chalmers University of Technology \\
and  University of G\"oteborg, S-412 96 G\"oteborg,
Sweden \\ }}
%
\end{center}
%
\begin{center}
{\bf Abstract} \\
\end{center}
\begin{quotation}
{\normalsize\noindent
The one--loop effective action for a slowly varying electromagnetic
field is computed at finite temperature and density
using a real-time formalism.
We   discuss the gauge invariance of the result.
Corrections to the Debye mass from an electric field
are computed at high temperature and high density.
The effective coupling
constant, defined from a purely
electric weak--field
expansion,
 behaves at high temperature very differently from
the case of a magnetic field, and does not satisfy
the renormalization group equation. The issue of pair production in the
real--time formalism is discussed
and also its relevance for heavy--ion collisions.}
\end{quotation}
\hfill
\newpage
%
\normalsize
\setcounter{page}{1}
\section{Introduction}\label{intro}
%
The formulation of quantum field theory in an external field is interesting
because of the many applications where the background field is strong
\cite{GreinerMR85} and cannot be treated perturbatively.
Some relativistic examples are the magnetic field in
neutron stars, white dwarfs   and heavy--ion collisions
\cite{ShapiroT83}. More extreme situations are given by the
electroweak phase transition   and  cosmic strings
\cite{Vachaspati91}. Strong (colour)
electric fields furthermore appear in some
models
of hadronization in heavy-ion collisions (for some recent discussions see e.g.
\cite{Baur90,Ganguly93}). In most
of the examples above the fields exist in a thermal
heat bath or some non--equilibrium background which is very different from the
vacuum. In the case of hadronization in heavy-ion collisions it has actually
been argued (see e.g. Ref.\cite{Ganguly93}) that there is a time-interval
during which local
thermal equilibrium has been achieved but the external (colour) electric
field has not yet been depleted due to particle production. For such a
time-interval the results of the present paper apply.

In this letter the
one--loop effective action for a constant
(or slowly varying) electromagnetic field
\cite{Schwinger51} is generalized to finite temperature and chemical potential,
$(T,\mu)$. Or, to put it differently, the thermal effective action for a
constant magnetic field \cite{ElmforsPS93} is generalized to arbitrary constant
electromagnetic fields. The case of a pure magnetic field has been treated
earlier in \cite{Dittrich79,ChodosEO90,ElmforsPS93} and the incomplete
result of \cite{ChodosEO90} was corrected
in \cite{ElmforsPS93}.
The case of a general electromagnetic field has been
studied in \cite{CoxHY84,LoeweR92} and some corrections to
 the expressions for the effective action in
those papers are presented here.
For constant external fundamental fields there are various methods of
calculating the effective action to all orders in the field.
Schwinger calculated the effective action for a constant background {\it
field--strength} in QED, for which the gauge fields are non--constant
\cite{Schwinger51}. The proper--time method used in \cite{Schwinger51}
relies  on the fact that the solution of the quantum mechanical equations of
motion for a particle in the background field can be found explicitly,
and an extension to finite $(T,\mu)$ is possible using real--time
thermal propagators.
The case of a magnetic field has a rather
clear physical interpretation at finite
$(T,\mu)$ with the particles in the
heat bath occupying the time--independent Landau
levels. This situation has been studied in detail in \cite{ElmforsPS93}.
In the presence of a slowly varying electric field the particles
in equilibrium
screen the field over a distance determined by the Debye screening length.
  Schwinger's calculation for a general slowly varying electromagnetic field
can be extended to
finite $(T,\mu)$
in a rather straightforward manner without finding the
particle spectrum explicitly.
The constant field approximation
is satisfied if the relative gradient of the field is smaller
than any other scale, i.e.
$|\partial_\alpha F_{\mu\nu}|/|F_{\mu\nu}|\ll |eF_{\mu\nu}|^{1/2},
\ m_e,\ T$ or $n^{1/3}$,
where $n$ is the particle density. In this letter we consider only the
non--interacting $e^+e^-$--gas in a time--independent and slowly varying
background
where the conditions above are satisfied. For instance, the result applies to a
shallow potential well with an arbitrary constant
magnetic field and low density.
It has been strongly argued against the possibility of having
an electric field and a net charge density at equilibrium
\cite{Muller91}. We want to stress that this may be true for a strictly
constant electric field, but it is perfectly physical to have a slowly varying
field at thermal equilibrium and consider an expansion in derivatives in the
field,
at least for a stable system.
A deeper analysis is
required to determine whether the result can be used for more general
situations.
  We shall limit the discussion here to
the thermal corrections since the vacuum part is easily added
when needed.
%
\section{One--loop effective action}\label{ea}
Schwinger's equation for the one--loop effective action, generalized to finite
temperature and density, can be written as
\begin{equation}\label{dGdm}
        \frac{\partial \Gamma[A]}{\partial m}=i{\rm Tr}\langle
x|\frac{1}{\pi\!\!\!/-m+i\epsilon}
        -f_F(p_0,A_0)\left(\frac{1}{\pi\!\!\!/-m+i\epsilon}-
        \frac{1}{\pi\!\!\!/-m-i\epsilon}\right)|x\rangle\ ,
\end{equation}
where ${\rm Tr}$ is the trace over spin and $x$,
$\pi\!\!\!/=\gamma^\mu(p_\mu-eA_\mu)$, and $f_F(p_0,A_0)$ is the thermal
distribution function.  Under the gauge transformation $A_\mu(x)\mapsto
A_\mu(x)+\partial_\mu\Lambda(x)$,
the zero temperature part of Eq.(\ref{dGdm}) is transformed to
\begin{equation}\label{gtrfzero}
        i{\rm Tr} \langle x|e^{-ie\Lambda(x)}\frac{1}{\pi\!\!\!/-m+i\epsilon}
        e^{ie\Lambda(x)}|x\rangle\ ,
\end{equation}
and is thus gauge invariant.
Since the thermal part of Eq.(\ref{dGdm}) is not invariant under a general
gauge transformation we have to explain the apparent problem of gauge
dependence. The density matrix for the whole system of fermions and the
electromagnetic field is given by
\begin{equation}
\label{rhowhole}
        \rho_{QED}=\frac{\exp[-\beta\int d^3x (T^{00}_{QED}(x)-
        \mu e\overline{\Psi}\gamma^0\Psi)]}
        {{{\rm tr}} \exp[-\beta\int d^3x (T^{00}_{QED}(x)-
        \mu e\overline{\Psi}\gamma^0\Psi)]}\ ,
\end{equation}
where $T^{\mu\nu}_{QED}$ is the energy--momentum tensor
\begin{equation}
\label{Tmn}
        T^{\mu\nu}_{QED}=F^{\mu\alpha}F_\alpha^{~~\nu}+
        (\partial_\alpha F^{\mu\alpha}) A^\nu
        +i \overline{\Psi}\gamma^\mu\partial^\nu\Psi
        +\frac{1}{4} g^{\mu\nu} F_{\alpha\beta}F^{\alpha\beta}-
        g^{\mu\nu}\overline{\Psi}(i \partial\!\!\!/-eA\!\!\!/ -m)\Psi\ .
\end{equation}
One can easily verify that $T^{\mu\nu}_{QED}$ is gauge invariant when Maxwell's
equation $\partial_\alpha F^{\alpha\mu}=e\overline{\Psi}\gamma^\mu\Psi$ is
satisfied. It is, however, common to consider the fermions in the background
field   $F^{\mu\nu}$ separately and only use the fermionic part of
$\rho_{QED}$ to define the density matrix. The energy defining $\rho_{e^+e^-}$,
\begin{equation}\label{Ham}
        P_0=\int\!d\,^3x T^{00}_{e^+e^-}=
        \int\!d\,^3x[\overline{\Psi}(i \gamma_i\partial_i-e\gamma_i A_i
         +m)\Psi+e\overline{\Psi}\gamma_0 A_0\Psi]
        =\int\!d\,^3x
        \overline{\Psi}i\gamma_0 \partial_0\Psi\ ,
\end{equation}
is then gauge dependent. It is not even a conserved quantity in a general
gauge, in contrast to the total energy of the particles and the field.
The equilibrium $\rho_{e^+e^-}$ can also be defined as the state with maximal
entropy for a given expectation value of $P_0$. It is then clear that the
separation of the background field and the fermions is only meaningful in a
gauge where $P_0$ is a conserved quantity.
Only for
time independent $A_\mu(x)$ is $P_0$ separately
conserved (assuming $F_{\mu\nu}$ to be constant)
and that determines the gauge we
have to use, namely $\partial_0 A_\mu(x)=0$,
up to time independent gauge transformations which,
anyway, leave
the final result invariant.\footnote{
Notice that such a gauge was not used in \cite{CoxHY84}.}
 To be more precise, $\Lambda(x)$
can a have time dependence of the form
$\Lambda(x)= c\cdot t + \lambda({\bf x})$, where $c$ is a constant
and $\lambda({\bf x})$ is an arbitrary function.
This transformation shifts the potential $A_0$ by the
constant $c$. Such a constant is actually   absorbed into the definition of
the chemical potential, which is not an independent physical parameter but is
to be determined by a given charge density.
Only the difference $\mu-eA_0$ is physically meaningful for $\rho_{e^+e^-}$.
Note that this discussion is equally relevant when $F_{\mu\nu}=0$
but then one normally puts $A_\mu=0$.

To find
thermal one--particle distribution function $f_F(p_0,A_0)$ we argue that at
high
temperature it must be reduced to the classical Boltzmann distribution for
electrons and positrons, $\exp[-\beta(\sqrt{{\bf p}^2+m^2}\pm eA_0\mp\mu)]$,
and
it is the sign of $(p_0-eA_0)$ that distinguishes between particles and
anti--particles. We therefore write
\begin{equation}\label{fF}
        f_F(p_0,A_0)=\frac{\theta(p_0-eA_0)}{e^{\beta(p_0-\mu)}+1}+
        \frac{\theta(-p_0+eA_0)}{e^{\beta(-p_0+\mu)}+1}\ .
\end{equation}
The distribution function does not have
to be chosen to represent an equilibrium distribution.
Other choices may be appropriate when the electric field
drives the system out of equilibrium.
It has been emphasized in the literature (see e.g.
Ref.\cite{Muller91}) that the distribution in Eq.(\ref{fF})
has a non--trivial limit when $T\rightarrow 0$, and describes a Dirac sea
filled up to the energy $\mu$.

{}From Eq.(\ref{dGdm})
one can follow the calculations in \cite{Schwinger51} to obtain
the thermal part of the effective action. The only difference
is that the trace should be taken in the basis $|p_0,{\bf x}\rangle$.
We find
\begin{eqnarray}\label{genform}
        \frac{\partial {\cal L}^{\beta,\mu}_{\rm eff}(x)}{\partial m^2}&=& -
        \frac{1}{2\pi^{3/2}}\int_{-\infty}^\infty\frac{dp_0}{2\pi}
        f_F(p_0,A_0){\rm Im}\Biggl\{\int_0^\infty ds
        e^2ab\cot(esa)\coth(esb)\nonumber \\*
        && \times (h(s)-i\epsilon)^{-1/2}
        \exp\left[-i(m^2-i\epsilon)s
        +i\frac{(p_0-eA_0)^2}{h(s)-i\epsilon}-i\frac{\pi}{4}\right]\Biggr\}\ ,
\end{eqnarray}
where
\begin{eqnarray}\label{symbols}
        h(s)& = &(eF\coth eFs)_{00}\ ,\nonumber \\*
        F_{{\mu\nu}}& = &\partial_\mu A_\nu-\partial_\nu A_\mu\ ,\nonumber \\*
        a^2-b^2&=& B^2-E^2\equiv 2{\cal F}
        =\frac{1}{2}F_{{\mu\nu}} F^{{\mu\nu}}\ ,\nonumber \\*
        ab&=& {\bf E}\cdot{\bf B}\equiv{\cal G}
        =\frac{1}{4}F^*_{{\mu\nu}}F^{{\mu\nu}}\ , \nonumber \\*
        B&=&|{\bf B}|\ ,\quad E=|{\bf E}|\ .
\end{eqnarray}
We have added   $-i\epsilon$ to $h(s)$ in Eq.(\ref{genform}) in order to get
the
correct value using the principle branch when $h(s)<0$. It is determined by the
$x_0$--integration of Schwinger's formula. From the $\cot(esa)$ factor
we find that the $s$--integral goes through a number of poles on
the real axis, $s=k\pi/ea$,
which were not apparent in the original expression. In
the case of a pure magnetic field it has been shown
\cite{ElmforsPS93} that the poles are
absent if we only include a finite number of Landau levels, and to get the
correct result after summing over all Landau levels the
$s$--integration contour has to go slightly below the real axis.
We assume that the same prescription is valid even for non--zero $E$ field
since the poles are related to the $B$ field.
The expression in Eq.(\ref{genform}) can be directly integrated
with respect to $m^2$. The singularity at $s=0$ should be
cured by subtracting the $F_{\mu\nu}=0$ part or using a $\zeta$--function
regularization as in \cite{ElmforsPS93} (or using Eq.(\ref{sinint})).
We  notice that in the limit $T\rightarrow \infty$,
and $f_F(p_0,A_0)\rightarrow 1/2$, Eq.(\ref{genform})
equals the negative of the real part of the vacuum contribution
up to divergent terms which are only quadratic in the field.

 It is possible to find an
explicit expression for $h(s)$ using the following observations. The
characteristic equation for the fieldstrength tensor
$F \equiv F_{\mu}^{~\nu}$
is \begin{equation}\label{char}
        \det(\lambda{\leavevmode\hbox{\small1\kern-3.3pt\normalsize1}}-F)=
\lambda^4+2{\cal F} \lambda^2-{\cal G}^2=0\ ,
\end{equation}
with the eigenvalues
\begin{equation}\label{eigenvals}
        \lambda=\pm i (\sqrt{{\cal F}^2+
         {\cal G}^2}+{\cal F})^{1/2}\ ,\quad
        \pm  (\sqrt{{\cal F}^2+ {\cal G}^2}-{\cal F})^{1/2}\ .
\end{equation}
We know that $(F\coth F)$ has a formal
power series expansion involving only even
powers of $F$. The Cayley--Hamilton theorem states that $F$ satisfies its own
characteristic equation, which can be used to reduce the powers of $F$ down to
at most $F^2$, so the matrix structure must be
\begin{equation}\label{fcothf}
        F\coth F=\gamma
        {\leavevmode\hbox{\small1\kern-3.3pt\normalsize1}}+\delta F^2\ .
\end{equation}
Taking the trace and determinant on both sides of Eq.(\ref{fcothf}) gives two
equations to determine $\gamma$ and $\delta$.
We find
\begin{eqnarray}\label{alpha}
        && h(s)=\nonumber\\[2mm]
        &&\frac{e}{2}
        \left(1-\frac{(E^2+B^2)}{2\sqrt{{\cal F}^2+{\cal G}^2}}\right)
        (\sqrt{{\cal F}^2+ {\cal G}^2}+{\cal F})^{1/2}\cot
        \big[es(\sqrt{{\cal F}^2+
        {\cal G}^2}+{\cal F})^{1/2}\big]\nonumber\\[2mm]
        \!\!\!\!&+&\frac{e}{2}\left(1+\frac{(E^2+B^2)}
        {2\sqrt{{\cal F}^2+{\cal G}^2}}\right)
        (\sqrt{{\cal F}^2+{\cal G}^2}-{\cal F})^{1/2}\coth
        \big[es(\sqrt{{\cal F}^2+{\cal G}^2}-
        {\cal F})^{1/2}\bigg]\ .
        ~~~~~
\end{eqnarray}
The relation $s\cdot h(s)\geq 0$ is thus not valid for
general electromagnetic field, contrary to the claim in \cite{LoeweR92}.
The quantity $\sum_{\mu=1,2,3}[\ln(\sinh(eFs)/eFs)]_\mu^{\ \mu}$,
appearing in the effective action
in \cite{CoxHY84}, can be  calculated using  the same method,
and we do not agree on their result either.
The standard way to proceed from Eq.(\ref{genform})
is to deform the $s$--contour to the
imaginary axis. Here, it is not always   useful,
except in special cases, since $h(s)$ has zeros on the
imaginary axis which give  essential
singularities from the exponential. On the other hand, the same
problem occurs also on the real axis.
To be more specific we can consider the
following special cases
\begin{equation}\label{fall}
\begin{array}{lrl}
        {\rm\bf I.}& {\bf E}\parallel {\bf B} :& h(s)=eE\coth esE
        \ ,\nonumber\\[2mm]
        {\rm\bf II.}& {\bf E}\perp {\bf B}, \ E>B:&
          {\displaystyle h(s)=\frac{e}{E^2-B^2}\left(E^2\sqrt{E^2-B^2}
          \coth es\sqrt{E^2-B^2}-\frac{B^2}{es}\right)}\ ,\nonumber\\[4mm]
        {\rm\bf III.}& {\bf E}\perp {\bf B},\ B>E:&
          {\displaystyle h(s)=\frac{e}{B^2-E^2}\left(\frac{B^2}{es}-
          E^2\sqrt{B^2-E^2}\cot es\sqrt{B^2-E^2}\right)}\ .
\end{array}
\end{equation}
In the cases {\bf I} and {\bf II}
$h(s)$ has no zeros for real $s$ so one
kan keep the contour as in Eq.(\ref{genform}). A special case
of {\bf III} is when $E\rightarrow 0$
and we get back the result in \cite{ElmforsPS93}
by deforming the contour to the imaginary axis. Another
 interesting special case
of ${\bf E}\perp {\bf B}$ is when $E=B$
in which case we  have $a=b=0,\ h(s)=1/s,$ just like in absence
of the external field, but with a potential $A_0$. We then find
\begin{eqnarray}\label{fnoll}
        &&{\cal L}^{\beta,\mu}_{\rm eff}(F_{\mu\nu}=0,A_0)=
        {\cal L}^{\beta,\mu}_{\rm eff}({\bf E}\perp{\bf B},E=B)
        \nonumber \\*
        &&=\frac{1}{3\pi^2}\int dp_0 f_F(p_0,A_0)
        \theta((p_0-eA_0)^2-m^2)((p_0-eA_0)^2-m^2)^{3/2}\ ,
\end{eqnarray}
a result similar to the absence of quantum corrections in a
propagating plane wave at zero temperature \cite{Schwinger51}.
Notice that, because of the lack of Lorentz invariance,
this observation is non--trivial since ${\cal L}^{\beta,\mu}_{\rm eff}$ does
not
only depend on ${\cal F}$ and ${\cal G}$.
To obtain Eq.(\ref{fnoll}) we have used dimensional regularization
as in \cite{ElmforsPS93} and the relation
\begin{equation}\label{sinint}
        \int_0^\infty ds\, s^\nu\sin(as-\frac{\pi}{4})
        =\frac{\Gamma(\nu+1)}{|a|^{\nu+1}}\left\{
        \begin{array}{rl} \sin\frac{\pi}{2}
        (\nu+\frac{1}{2}) &\ {\rm if }\ a>0\ ,\\
                -\sin\frac{\pi}{2}(\nu+\frac{3}{2}) &\ {\rm if }\ a<0\ .
        \end{array}\right.
\end{equation}
%
\section{Weak $E$--field expansion}\label{weak}
The $\theta$--function in Eq.(\ref{fnoll}) arises from the non--analytic
behaviour of Eq.(\ref{sinint}). If we take $B=0$ and expand
the integrand of Eq.(\ref{genform}) in powers of $E$, using
Eq.(\ref{fall}), we find that the $s$--integral can be performed
term by term using Eq.(\ref{sinint}), resulting in the same
$\theta$--functions. However, the remaining $p_0$--integral
becomes infra--red divergent at $(p_0-eA_0)^2=m^2$. This problem
occurs already at ${\cal O}(E^2)$ in contrast to the case
of a $B$--field \cite{ElmforsPS93} where the ${\cal O}(B^2)$
term can be calculated by a direct expansion of the integrand.
It is not {\it a priori} clear that a power expansion
in $E^2$ exists, and it is also plausible that the expansion is
only asymptotic just like at zero temperature.
Such an expansion would still be useful in many
physical situations where the field is not too strong.
To find the leading weak field expansion we have to be
more careful than simply expanding the integrand.
Using the notation from Eqs.(\ref{fall},{\ref{fnoll}), we derive
from Eq.(\ref{genform})
\begin{eqnarray}\label{LE}
        {\cal L}^{\beta,\mu}_{\rm eff}(E)-{\cal L}^{\beta,\mu}_{\rm eff}(0)&=&
        -\frac{1}{4\pi^{5/2}}\int_{-\infty}^\infty d\omega f_F(\omega+eA_0)
        \int_0^\infty\frac{ds}{s^{5/2}}\nonumber \\*
        &&\Biggl\{\left[
        (h(s) s)^{1/2}\cos\left(\omega^2 s
        \frac{1-h(s) s}{h(s) s}\right)-1\right]
        \cos\left((\omega^2-m^2)s-\frac{\pi}{4}\right)\nonumber \\*
        &&-(h(s) s)^{1/2}\sin\left(\omega^2 s
        \frac{1-h(s) s}{h(s) s}\right)
        \sin\left((\omega^2-m^2)s-\frac{\pi}{4}\right)\Biggr\}\nonumber \\*
        &\equiv&{\cal L}_1(E)+{\cal L}_2(E)\ ,
\end{eqnarray}
where $\omega=p_0-eA_0$. In the first part of the curly
bracket in Eq.(\ref{LE}), called ${\cal L}_1$, we can use the expansion
$h(s) s\simeq 1+(seE)^2/3$ to find the finite
$E^2$ term
\begin{equation}\label{L1}
        {\cal L}_1(E)\simeq-\frac{(eE)^2}{24\pi^2}\int_{-\infty}^\infty
        \frac{d\omega}{(\omega^2-m^2)^{1/2}}
        f_F(\omega+eA_0)\theta(\omega^2-m^2)\ .
\end{equation}
The infra--red problems arise when trying to expand ${\cal L}_2$ in $E$.
After introducing a new integration variable $x=\omega^2$, and
performing a partial integration with respect to $x$,
we get
\begin{eqnarray}\label{L2}
        {\cal L}_2(E)&=&\frac{1}{8\pi^{5/2}}\int_0^\infty
        dx \int_0^\infty\frac{ds}{s^{7/2}}
        (h(s) s)^{1/2}\cos(xs-m^2s-\frac{\pi}{4})\nonumber \\*
        &&\frac{d}{dx}\left[
        \frac{f_F(\omega+eA_0)+f_F(-\omega+eA_0)}{\omega}
        \sin\left(xs\frac{1-h(s) s}{h(s) s}\right)\right]\
\end{eqnarray}
which can be expanded to ${\cal O}(E^2)$. To this order in the field we then
have that
\begin{equation}
\label{Last1}
        {\cal L}^{\beta,\mu}_{\rm eff}(E)-{\cal L}^{\beta,\mu}_{\rm eff}(0)
        = \frac{(eE)^2}{24\pi^2}\int_{m}^\infty
        \frac{d\omega\,\omega}{(\omega^2-m^2)^{1/2}}\frac{d}{d\omega}
        \left(\frac{1}{e^{\beta (\omega + A_{0} - \mu )} + 1}
+ \frac{1}{e^{\beta (\omega - A_{0} + \mu )} + 1}
\right)\ .
\end{equation}
Eq.(\ref{Last1}) leads to
\begin{equation}
        {\cal L}^{\beta,\mu}_{\rm eff}(E) =
        {\cal L}^{\beta,\mu}_{\rm eff}(0)-
        \frac{(eE)^2}{24\pi^2}
        \frac{|\mu - eA_{0}|}{\sqrt{(\mu - eA_{0})^2 - m^2}}
        \,\theta \left( |\mu - eA_{0}| - m \right) \ ,
\end{equation}
if $T=0$ and
\begin{equation}
{\cal L}^{\beta,\mu}_{\rm eff}(E) =
        {\cal L}^{\beta,\mu}_{\rm eff}(0)-
        \frac{(eE)^2}{24\pi^2}\left(1 -
        2\pi (\frac{m}{T})^{2} \sum_{l=0}^{\infty}
        \frac{1}{\left((m/T)^{2} + (2l+1)^{2}\pi ^{2}\right)^{3/2}}\right)\ ,
\end{equation}
if $\mu - eA_{0} = 0$.
The Debye mass can be extracted as the second derivative of
${\cal L}^{\beta,\mu}_{\rm eff}(E)$ with respect to $A_0$.
{}From the zero field part in Eq.(\ref{fnoll}) we get
$m^2_\gamma(\mu)\simeq e^2(\mu-eA_0)^2/\pi^2$  and
$m^2_\gamma(T)\simeq e^2 T^2/3$ in the high density and
temperature limit.  This agrees with
\cite{Weldon82,AltherrK92} but not with \cite{LoeweR92}.
Corrections from finite $E$--field, to lowest order in the field, can also be
calculated making use of
Eq.(\ref{Last1})
and  we find for high temperature and high
density, respectively
\begin{eqnarray}
\label{Debyemass}
        m^2_\gamma(\mu,E) &=&
        m^2_\gamma(\mu)
        - \frac{\alpha}{2\pi}\frac{(eE)^2}{m^2}\frac{m^4}{(\mu-eA_0)^4}\left(1+
{\cal
O}\left((\frac{m}{\mu-eA_0 })^{2}\right)\right)
        \ ,\nonumber\\
        m^2_\gamma(T,E) &=&
        m^2_\gamma(T)-\frac{\alpha}{2\pi}\frac{(eE)^2}{m^2}\frac{31}{4\pi
^{2}}\zeta
(5)
\frac{m^4}{T^4}\left(1 +{\cal O}\left((\frac{m}{T})^{2}\right)\right)\ ,
\end{eqnarray}
where $\zeta (n)$ is the Riemann zeta-function and $\alpha = e^{2}/4\pi $ is
the fine--structure constant.
We notice that the $E$--field tends to decrease the
screening mass.

In analogy with \cite{ChodosEA88,ElmforsPS93} we can define an effective
fine structure constant by
\begin{equation}\label{aqed}
        \frac{1}{\alpha (T,\mu)}=
        \frac{1}{\alpha }+\frac{1}{\alpha E}
        \left.\frac{\partial {\cal L}^{\beta,\mu}_{\rm eff}(E)}
        {\partial E}\right|_{E\rightarrow 0}\ .
\end{equation}
In the high density or temperature limit we then have that
$\alpha (T,\mu)\rightarrow\alpha /(1-\alpha /3\pi)$,
showing a completely different behaviour from the $\alpha $
defined using a magnetic field \cite{ElmforsPS93}, which
satisfies a zero temperature renormalization group
equation.
%
\section{Pair production}\label{pair}
One issue that has been discussed in the literature is pair production at
finite $T$ \cite{CoxHY84,LoeweR92}, which we find to be absent
 from the thermal part of
the one--loop contribution, in the real--time formalism. This can
be seen directly from Eq.(\ref{dGdm}) in which the thermal part is manifestly
real.
It is illustrative to compare with the standard calculation of
the effective potential $V(\phi)$ for a spontaneously broken
$\lambda\phi^4$--model
where one may find an imaginary part when the effective mass,
$M^2(\phi)=\lambda\phi^2/2-m^2$, is negative.
The expression
\begin{equation}\label{dVdphi}
        \frac{dV(\phi)}{d\phi}=\frac{\lambda\phi}{2}
        \int\frac{d\,^4p}{(2\pi)^4}f(p_0)
        \left(\frac{i}{p^2-M^2+i\epsilon}-
        \frac{i}{p^2-M^2-i\epsilon}\right)\ ,
\end{equation}
is obviously real for any real $M^2$. An imaginary part of $V(\phi)$
is found when first calculating Eq.(\ref{dVdphi}) for a positive
$M^2$ in the limit $\epsilon\rightarrow 0$ and then performing an (ambiguous)
analytic continuation in $M^2$.
This new function corresponds to, for negative $M^2$,
a $p_0$--integration contour which does not follow the
real axis but goes above (below) the poles for negative (positive)
$p_0$, or vice versa.
The procedure of obtaining an imaginary part in this way is thus
both ambiguous and does not correspond to the $\epsilon$--prescription
in Eq.(\ref{dVdphi}). The conclusion we can draw from the example above is that
the standard real-time calculation of the one--loop
effective potential gives no imaginary part.
One should, however, remember that the standard real--time
rules are usually derived under the assumptions
of certain factorization properties (see \cite{Pearson93} for a discussion)
 that may
not be fulfilled in presence of unstable modes. An imaginary part can be
obtained in the imaginary--time formalism, but it is not clear that a
consistent calculation of the pair production rate can be performed assuming
equilibrium.

We conclude with a remark on heavy quark production in
ultra--relativistic nuclear collisions, showing the physical importance of pair
production. This example serves as a motivation for finding a solution to the
problem with the imaginary
part. If  ${\cal R}_{g \rightarrow
q\bar{q}}$ is the production  rate of thermal gluon decay into
quark-antiquarks one finds e.g. that \cite{altherr&seibert93}
\begin{equation}
\label{altherreq}
\frac{{\cal R}_{g \rightarrow q\bar{q}}}{T^{4}}  \simeq 0.01~~,
\end{equation}
for quark masses $m$ such that $m/T \leq 1$. Taking the
imaginary part of the effective action
${\rm Im }\, {\cal L}^{\beta,\mu}_{eff}(E)$ found in
\cite{LoeweR92,Ganguly93} seriously, which we
question in our paper, we would find for $T \gg  m$ and $\mu = 0$ a rate
 \begin{equation}
 \frac{ {\rm Im }\,{\cal L}^{\beta,\mu}_{eff}(E)}{T^{4}}
\simeq \left(\frac{0.3}{T_{GeV}} \right)^{2}~~~,
\end{equation}
which can be as important as perturbative production rates. Here we
have used that $eE \simeq 1\, GeV^2$. It turns out that perturbative
production of heavy quark-antiquarks due to the time-variation in the
background electric field can be as large as the thermal gluon decay rate
Eq.(\ref{altherreq}). Since quark-antiquarks are spontaneously produced
by Schwinger's mechanism, the background electric field will be depleted, as
mentioned in the introduction.
The time--scale, $t_{d}$, for the depletion of the electric field we
estimate using Schwinger's expression for the rate of pair production,
$w_{s}$, and  $m t_{d}w_{s} \simeq E^2 /2$. Approximatively, this amounts to
assuming an
exponential decay of the background electric field, i.e. $eE(t) \simeq
eE \exp(-t/t_{d})$. The explicit time--dependence in the electric field
leads to pair production already at the one--loop level with a rate
${\cal R}_{E}$. A straightforward calculation leads to the result
\begin{equation}
\frac{{\cal R}_{E}}{T^{4}} \simeq
 \left(\frac{0.2}{T_{GeV}} \right)^{4}~~~,
\end{equation}
where we have used $\alpha _{s} = 0.32$ and considered a quark mass
$m = 0.2\, GeV$. If the pair-produced particles are assumed to be
in thermal equilibrium, one can expect a decrease of this
production rate from Pauli blocking
factors.
\vskip 1 cm
\noindent{\bf Acknowledgement}
\\
We wish to thank D. Persson and P. Liljenberg for close
collaboration on related topics that initiated this work.
One of the authors (P.E.)
 have also appreciated the discussions and hospitality
of the staff and visitors at
ENSLAPP (Annecy), and the financial support from the Swedish
Institute.
%

\end{document}